\documentclass[prl,aps,showpacs,twocolumn,superscriptaddress,floatfix]{revtex4-1}

\usepackage{epsfig}
\usepackage{amsmath}
\usepackage{amssymb}
\usepackage{amsfonts}
\usepackage{mathptmx}
\usepackage{eucal}
\usepackage{bm}
\usepackage{color}
\usepackage[colorlinks,linkcolor=blue,citecolor=blue]{hyperref}

\usepackage{epstopdf}

\usepackage{graphicx}

\usepackage{natbib}

\begin{document}

\title{Manifestation of a spin-splitting field in a thermally-biased Josephson junction.}
\author{F. S. Bergeret}
\email{sebastian\_bergeret@ehu.es}
\affiliation{Centro de F\'{i}sica de Materiales (CFM-MPC), Centro
Mixto CSIC-UPV/EHU, Manuel de Lardizabal 4, E-20018 San
Sebasti\'{a}n, Spain}
\affiliation{Donostia International Physics Center (DIPC), Manuel
de Lardizabal 5, E-20018 San Sebasti\'{a}n, Spain}
\author{F. Giazotto}
\email{giazotto@sns.it}
\affiliation{NEST, Instituto Nanoscienze-CNR and Scuola Normale Superiore, I-56127 Pisa, Italy}

\begin{abstract}
We investigate the behavior of a Josephson junction consisting of a ferromagnetic insulator-superconductor (FI-S)  bilayer tunnel-coupled to a superconducting electrode.  We show that the  Josephson coupling in the structure is strenghtened by the presence of the spin-splitting  field induced in the FI-S bilayer.  
Such  strenghtening   manifests itself  as an increase  of the critical current $I_c$ with  the amplitude of the exchange field.  
Furthermore, the effect can be strongly enhanced if the junction is taken out of equilibrium by a temperature bias. 
We propose a realistic setup to assess experimentally the magnitude of the induced exchange field, and predict a drastic deviation  of  the $I_c(T)$ curve ($T$ is the temperature) with respect to equilibrium.  
\end{abstract}

\maketitle

The interplay between superconductivity and ferromagnetism  in superconductor-ferromagnet (S-F) hybrids exhibits a large variety of effects studied along the last  years \cite{BVErmp,Buzdin2005}. 
 Experimental research  mainly focuses on the control of the $0-\pi$ transition in S-F-S junctions \cite{Buzdin1982,Ryazanov2001} (S-F-S) and on the creation, detection  and manipulation of triplet correlations in S-F hybrids \cite{BVE2001b,Kaizer06,BlamireScience,Birge,Aarts}.  
 From a fundamental point of view, the key phenomenon for the understanding of these effects  is  the \emph{proximity} effect in S-F hybrids, and  how the interplay between superconducting and magnetic correlations affect their thermodynamic and transport properties. 
 
\begin{figure}[t!]
\begin{center}
\includegraphics[width=\columnwidth]{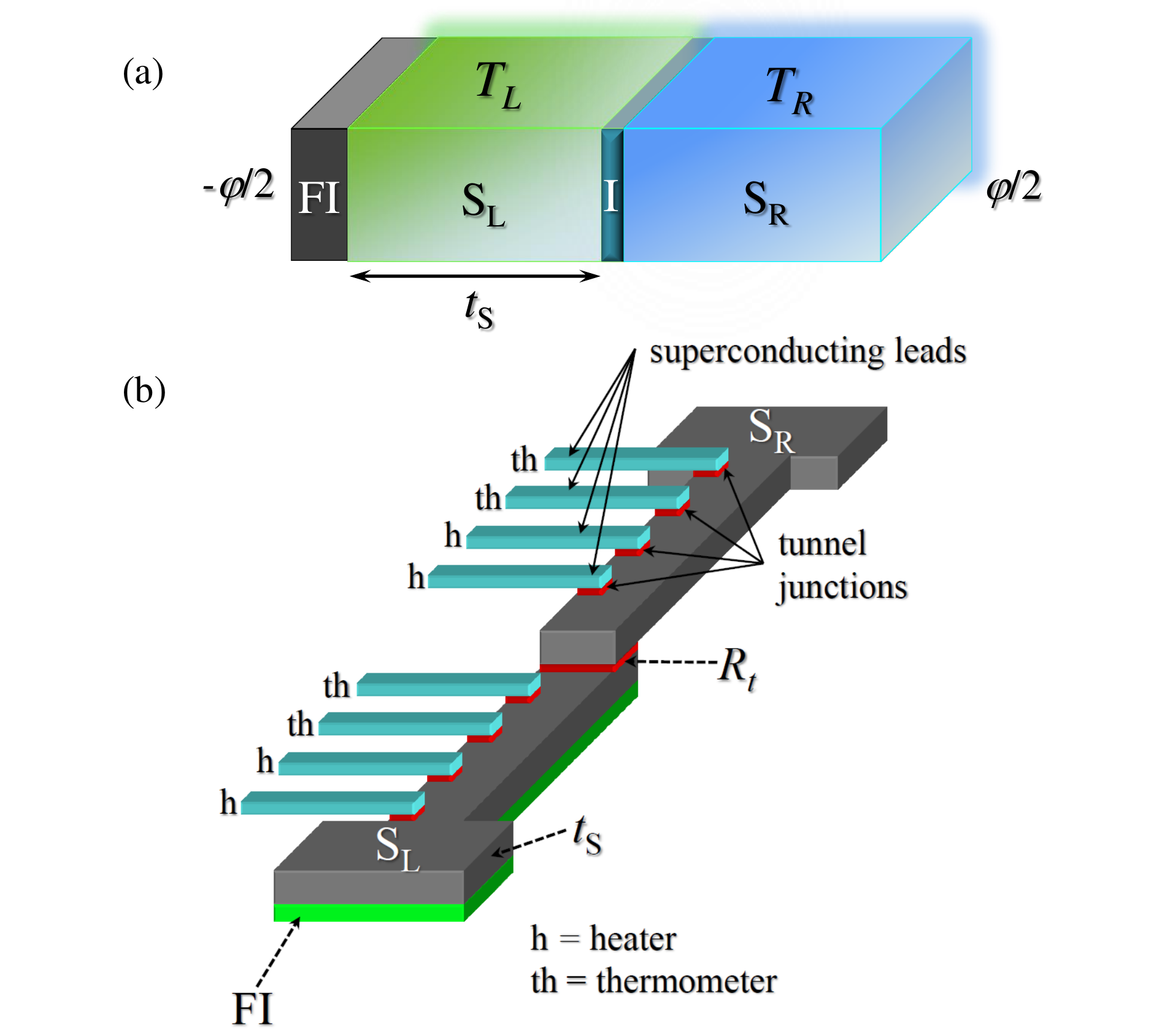}\vspace{-1mm}
\caption{(Color online) (a) Scheme of the FI-S-I-S Josephson tunnel junction considered in this paper. $T_L$ and $T_R$ indicate the temperature in the left (S$_{\text{L}}$) and right (S$_{\text{R}}$) superconductor, respectively, I stands for a conventional insulator whereas $\varphi$ is the macroscopic quantum phase difference over the junction. $t_{\text{S}}$ denotes the thickness of the S$_{\text{L}}$ layer. 
(b) Sketch of a possible experimental setup. Additional superconducting leads tunnel-coupled to S$_{\text{L}}$ and S$_{\text{R}}$
 serve either as heaters (h) or thermometers (th),
and allow one to probe the effect of a spin-splitting field through measurement of the junction current vs voltage characteristics under conditions of a temperature bias, as discussed in the text. $R_t$ denotes the junction normal-state resistance.}
\label{fig1}
\end{center}
\end{figure}
While most of theoretical and experimental investigations on S-F structures deal mainly with the penetration of the superconducting order into the  ferromagnetic regions, it is also widely known that magnetic correlations can be induced in the superconductor via the \emph{inverse} proximity effect \cite{Sauls88a,Bergeret2004,Bergeret2005,Kapitulnik2009}.
If the ferromagnet is an insulator (FI), on the one hand superconducting correlation are weakly suppressed at the FI-S interface  and a finite exchange field, with an amplitude smaller than the superconducting gap $\Delta_0$,  is induced at the interface. Such exchange correlations penetrate into the bulk of S over distances of the order of the coherence length \cite{Sauls88a}. This results in a splitting of the density of states (DoS) of the  superconductor, as observed in a number of experiments \cite{Moodera90,Moodera08,Catelani2008,Catelani2011}.
 Yet, the spin-split DoS of a superconductor may lead to interesting effects such as, for instance, the absolute spin-valve effect \cite{Mersevey1994,Nazarov2002,GT2008}, the magneto-thermal Josephson valve \cite{GB2013a,GB2013b}, and the large enhancement of the Josephson coupling observed in F-S-I-S-F junctions  (I stands for a conventional insulator) when the magnetic configuration of the F layers is arranged in the antiparallel state \cite{BVE2001a,Robinson2010}.

In this Letter  we show that an enhancement of the Josephson effect between two tunnel-coupled superconductors S$_\text{L}$ and S$_\text{R}$ can also be achieved if a unique FI is attached to one of the S electrodes, for instance, the left lead, as shown schematically in  Fig. \ref{fig1}(a). 
 According to the discussion above, the presence of the FI splits the  DoS in the left superconductor. 
 In principle, the presence of the spin-splitting field causes a reduction of the superconducting gap ($\Delta_L$) in the left superconductor, and therefore at first glance one may think that, in turn,  the Josephson coupling is suppressed. 
However, we show that for low enough temperatures,  the presence of the exchange field  $h$ in  one of the two electrodes indeed enhances the critical current ($I_c$) with respect to its value at  $h=0$.  
This effect is further enhanced  by applying  a temperature bias across the junction. Furthermore,  by setting S$_\text{L}$ at $T_L$ and S$_\text{R}$ at $T_R$,  the temperature-dependent $I_c$ curves change drastically, showing  a sharp step when the energy gaps difference matches the exchange field.
A measurement of $I_c$ therefore allows  to assess the magnitude of the induced $h$.   
We discuss different realizations, and propose  
a realistic setup and materials combinations to demonstrate our predictions. 
 
\begin{figure}[t]
\begin{center}
\includegraphics[width=\columnwidth]{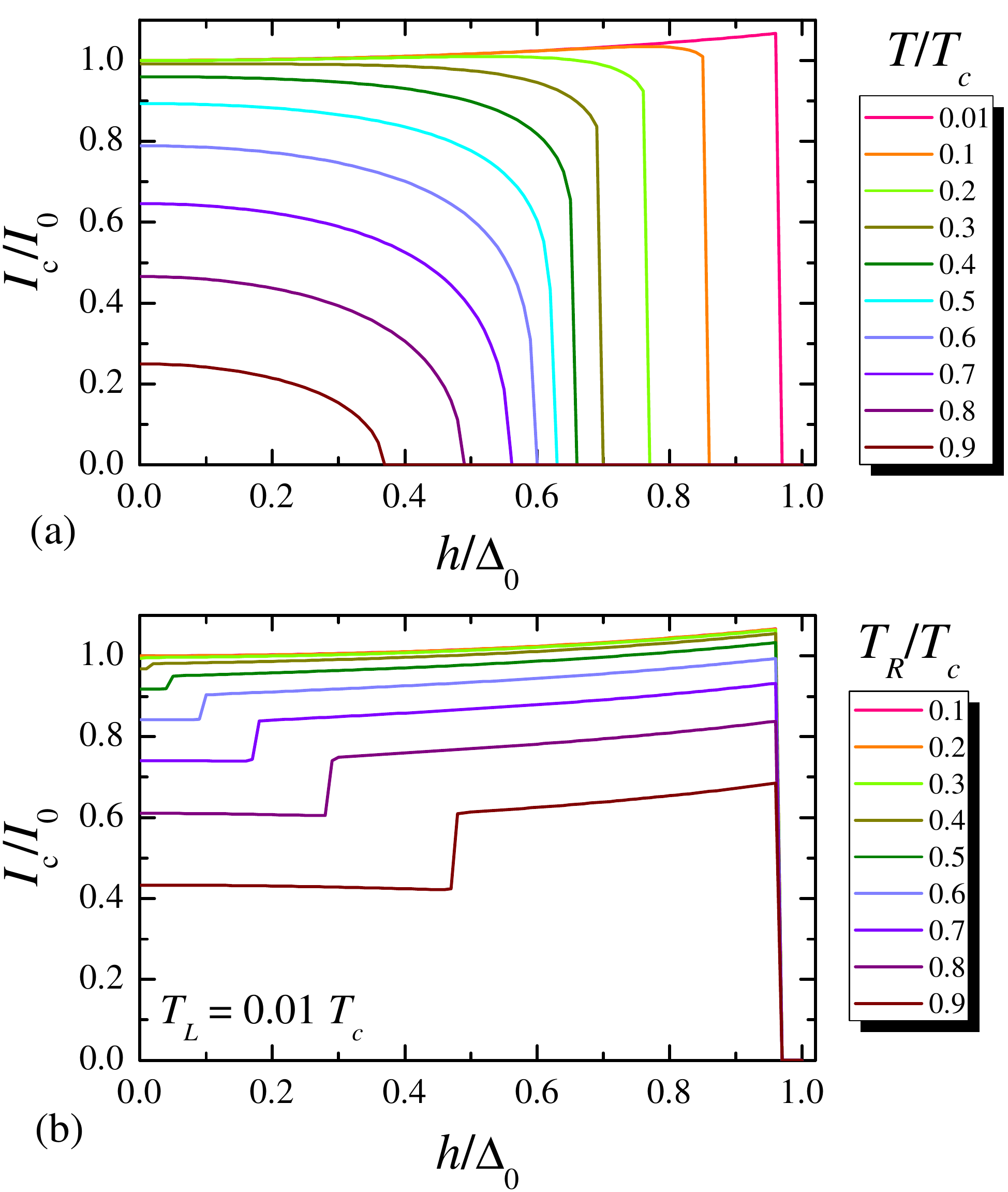}
\caption{(Color online) (a) Junction critical current $I_c$ versus exchange field $h$ calculated for several values of the temperature. Here we set $T_L=T_R=T$.
(b) Critical current $I_c$ versus $h$ calculated for different values of $T_R$ at $T_L=0.01T_c$. $\Delta_0$ denotes the zero-field, zero-temperature superconducting energy gap with critical temperature $T_c$. }\label{fig2}
\end{center}
\end{figure}

In order to understand the enhancement of the Josephson coupling by increasing the exchange field $h$ we provide here a simple physical picture that involves two mechanisms: 
On the one hand, the Josephson effect in the FI-S-I-S junction of Fig. \ref{fig1}(a) is stronger the larger the overlap of the condensates from the left an right electrodes is. 
This overlap is proportional to  the number of Cooper pairs with shared electrons between S$_\text{L}$ and S$_\text{R}$. 
By increasing the exchange field in the left side of the junction, it is energetically more favorable for the electrons with spin parallel to the field (spin-up) to be localized within S$_\text{L}$, whereas  spin-down electrons are preferably localized in S$_\text{R}$ where the exchange field is absent. 
This means that
the number of  Cooper pairs sharing  becomes \emph{larger}.  
On the other hand, the Josephson coupling is proportional to the amplitude of the condensate in each of the electrodes. Therefore by increasing $h$ or the temperature ($T$)  one expects a suppression of the order parameter in the electrodes. 
The behavior of the Josephson critical current as a function of $h$ and $T$ is therefore the result of these two competing   mechanisms. 
In the structure under consideration the exchange field acts only in S$_\text{L}$. 
For low enough temperatures, $\Delta_L(h,T)$ depends only weakly on $h$.  
Therefore the first mechanism dominates  and $I_c$ is enhanced by increasing $h$ [see Fig. \ref{fig2}(a)]. 
At large enough temperatures, $\Delta_L(h,T)$ is much more sensitive to the exchange field, and its faster suppression leads to a decrease of $I_c$ upon increasing $h$.  
We note that in S$_\text{R}$ the exchange field is absent, and therefore the suppression of $\Delta_R$ is caused only by the increase of the temperature. 
If we now keep S$_\text{L}$ at low temperature and  vary only the right electrode temperature ($T_R$), the first mechanism dominates for  any value of $T_R$, and the enhancement of $I_c$ by increasing $h$ can always be observed [{\it cf.} Fig.\ref{fig2}(b)]. This is a remarkable effect that we now analyze quantitatively in the following.

In order to compute the Josephson current through the junction sketched in Fig. \ref{fig1}(a)  we assume that the normal-state resistance of the tunneling barrier $R_t$ is much larger than the normal-state resistances of the junction electrodes. 
In such a case, the charge current through the junction  can be calculated from the  well known expression 
\begin{equation}
I=\frac{1}{32eR_t}\int  {\rm Tr}\left\{\tau_3\left[{ \bf G_L}(E),{ \bf G_R}(E)\right]^K\right\}dE\; ,
\label{current}
\end{equation}
where the matrix $\tau_3$ in Eq. (\ref{current}) is the third Pauli matrix in the particle/hole space, 
 the Green functions (GFs) ${\bf G_{L(R)}}$ are the bulk GFs matrices for the left and right electrodes, and $e$ is the electron charge. 
 They are 8$\times$8 matrices in the Keldysh-particle/hole-spin space with the structure
 \begin{equation}
{\bf G_{L(R)}}=\left( \begin{array}{cc}
\check G_{L(R)}^R&\check G_{L(R)}^K\\
0&\check G_{L(R)}^A
\end{array}
\right).\label{G}
\end{equation}
We assume that the junction is temperature biased so that the left (right) electrode is held at a constant and uniform temperature $T_{L(R)}$,
and $\varphi$ denotes the macroscopic phase difference between the superconducting electrodes.
 In this case the retarded (R) and advanced (A) components are 4$\times$4 matrices in particle/hole-spin space defined as 
\begin{equation}
\check G^{R(A)}_{L(R)}=\hat g_{L(R)}^{R(A)}\tau_3+\hat f_{L(R)}^{R(A)}( i\tau_2\cos(\varphi/2) \pm i\tau_1\sin(\varphi/2)),
 \end{equation}
where $\hat g$ and $\hat f$ are matrices in spin-space defined by 
$\hat g_{L(R)}^{R}= g_{-L(R)}^{R}\sigma_3+g_{+L(R)}^{R}\sigma_0$ and 
$\hat f_{L(R)}^{R}= f_{-L(R)}^{R}\sigma_3+f_{+L(R)}^{R}\sigma_0$. 
In principle the  functions in the left electrode may depend on the spatial coordinates. 
To simplify the problem we assume that the thickness $t_S$ of the S$_{\text{L}}$ electrode is smaller than the superconducting coherence length and hence the Green's functions are well approximated by spatially constant functions 
\begin{eqnarray}
\hat f_{\pm L(R)}^{R}&=&\frac{1}{2}\left[\frac{\Delta_{L(R)}}{\sqrt{(E+h+i\Gamma)^2-\Delta_{L(R)}^2}}\right.\nonumber\\
&&\pm\left. \frac{\Delta_{L(R)}}{\sqrt{(E-h+i\Gamma)^2-\Delta_{L(R)}^2}}\right]\;. 
\end{eqnarray}
$\hat g_{\pm L(R)}^R$ has a similar form by replacing $\Delta_{L(R)}$ in the numerators of the previous expressions with  $E\pm h$.
For the particular setup of  Fig. \ref{fig1}(a), the exchange field in the right electrode is set to zero ($h=0$) and therefore $g_{-L}=f_{-L}=0$ and $f_R=f_{+R}$. 
Notice that the gaps $\Delta_{L(R)}$ depend on the corresponding temperature $T_{L(R)}$ and exchange field, and have to be determined self-consistently.
The advanced GFs have the same form after replacing $i\Gamma\rightarrow -i\Gamma$. The latter parameter describes inelastic effects within the time relaxation approximation \cite{gamma}. Finally, the Keldysh component of the GF [Eq. (\ref{G})] is defined as
\begin{equation}
\check G_{L(R)}^K=(\check G_{L(R)}^R-\check G_{L(R)}^A)\tanh({E}/{2T_{L(R)}}).
\label{GK}
\end{equation}
By using Eqs. (\ref{G}-\ref{GK}) we can compute the electric current from Eq. (\ref{current}). 
In the absence of a voltage drop across the junction (i.e., $V=0$) the charge current equals the Josephson current,  $I_J=I_c\sin\varphi$, where the critical supercurrent  is given by the expression
\begin{eqnarray}
I_c&=&\frac{i}{8eR_t}\int  dE\left\{ \left[ f^R_{R}f^R_{+L}-f^A_{R}f^A_{+L}\right]\left[\tanh(\frac{E}{2T_R})+\tanh(\frac{E}{2T_L})\right]\right. \nonumber \\
 &+&\left. \left[f^R_{R}f^A_{+L}-f^A_{R}f^R_{+L}\right]\left[\tanh(\frac{E}{2T_R})-\tanh(\frac{E}{2T_L})\right]\right\}.
 \label{Ic}
\end{eqnarray}
The second line of the above expression corresponds to the contribution from out-of equilibrium conditions due to a temperature bias across the junction. It vanishes when both electrodes are held at the same temperature and, as we will see below, leads to important deviations of $I_c(T)$ from its equilibrium behavior. 

Before analyzing the most general case,  we first
 assume equilibrium,  i.e.,  $T_L=T_R=T$ and compute   the Josephson critical current as a function of the exchange field. This is shown  in Fig. \ref{fig2}(a). 
At low enough  temperatures, $I_c$ increases by increasing the exchange field. 
This is an unexpected result since the increase of the exchange field in the left electrode reduces the corresponding self-consistent  gap $\Delta_L$, and therefore at first glance this suppression might lead to a reduction of $I_c$. 
However,  this mechanism competes with the  Josephson coupling, which, within the simple physical picture given in the introduction, is enhanced thanks to the fact that the electrons of the Cooper pairs with spin projection parallel to the field $h$ prefer to be localized mainly in S$_\text{L}$ while those with antiparallel spin are mostly localized in S$_\text{R}$. 

To quantify the effect it is convenient to consider the limiting case $T\rightarrow 0$ so that the critical current [Eq. (\ref{Ic})] can be written as 
\begin{equation}
I_c^{eq}(T=0)=\frac{\Delta_0^2}{2eR_t}\int \frac{dE}{\sqrt{E^2+\Delta_0^2}}{\rm Re}\left[\frac{1}{\sqrt{(E+ih)^2+\Delta_0^2}}\right],
\end{equation}   
where $\Delta_0$ is the superconducting gap at $T=0$ and $h=0$.  
For small values of $h\ll\Delta_0$ one can expand this expression and find
\begin{equation}
I_c^{eq}(T=0)\approx \frac{\pi\Delta_0}{2eR_t}\left(1+\frac{1}{8}\frac{h^2}{\Delta_0^2}\right)\; ,
\end{equation}
which confirms the enhancement of $I_c$ upon increasing $h$. 
In the opposite limit,  i.e.,  $h\rightarrow \Delta_0$, numerical evaluation of the integral gives
 $eR_tI_c^{eq}(T=0,h=\Delta_0)\approx1.69\Delta_0$ which is larger then the expected value at $h=0$, i.e.,  $\pi\Delta_0/2$ \cite{Tinkham}. 
This contrasts  with what obtained for the critical current of  a  F-S-I-S-F structure with magnetizations in the F layers arranged in the  antiparallel configuration, which diverges as $h\rightarrow\Delta_0$ \cite{BVE2001a}.  
Therefore, although a larger effect can be achieved in a S-F-I-S-F (or FI-S-I-S-FI junction),  for practical purposes the setup of figure  Fig. \ref{fig1}(a) with just one single FI is much simpler, and the measurement of $I_c$ enhancement does not require control of magnetizations direction. Moreover, in our geometry   one can boost the supercurrent enhancement  by applying a temperature bias across the junction, as we shall discuss in the following.

If the  temperatures in the superconductors  are different ($T_L\neq T_R$), although each of the electrodes is in local steady-state equilibrium, the junction as a whole is  in an out-of-equilibrium condition.  
In such a situation, also the second line  in Eq. (\ref{Ic}) contributes to the amplitude of the critical current, and  leads to  new features in the dependence of $I_c$ on $h$, and on the temperature difference.  
For instance, one can hold  S$_\text{L}$ at some fixed $T_L$ and vary the temperature $T_R$ of S$_\text{R}$,  or vice versa. 
The critical current can be calculated numerically from Eq. (\ref{Ic}). 
These results are  shown in panel (b) of Fig. \ref{fig2} where we set $T_L=0.01 T_{c}$, and $T_R$ varies from $0.1T_{c}$ up to $0.9T_{c}$.  
It is clear that, for large values of the spin-splitting field, $I_c$ is larger than the one for $h=0$.
It is also remarkable that the effect is more pronounced the  larger is the temperature difference. 
Furthermore, the main enhancement occurs stepwise, and stems from the out-of-equilibrium contribution  to $I_c$ appearing in Eq. (\ref{Ic}). The latter is equivalent to the expression for $I_{J_1}(V,T)$, the term proportional to $\sin\varphi$, of a voltage-biased Josephson junction obtained several years ago in Refs.\cite{LO,Harris}. In our system the exchange field plays the role of the voltage bias and, in agreement with Refs. \cite{LO,Harris}, the jump takes place  at the value of $h$ for which the following condition is satisfied:
\begin{equation}
\label{condition}
|\Delta_R(T_R)-\Delta_L(T_L,h)|=h\; .
 \end{equation}
We stress that while $I_{J_1}(V,T)$ can be accessed experimentally through a measurement of the ac Josepshon effect in voltage-biased configuration, the experiment we purpose below requires only a rather simple dc measurement at $V=0$.

\begin{figure}[t]
\begin{center}
\includegraphics[width=\columnwidth]{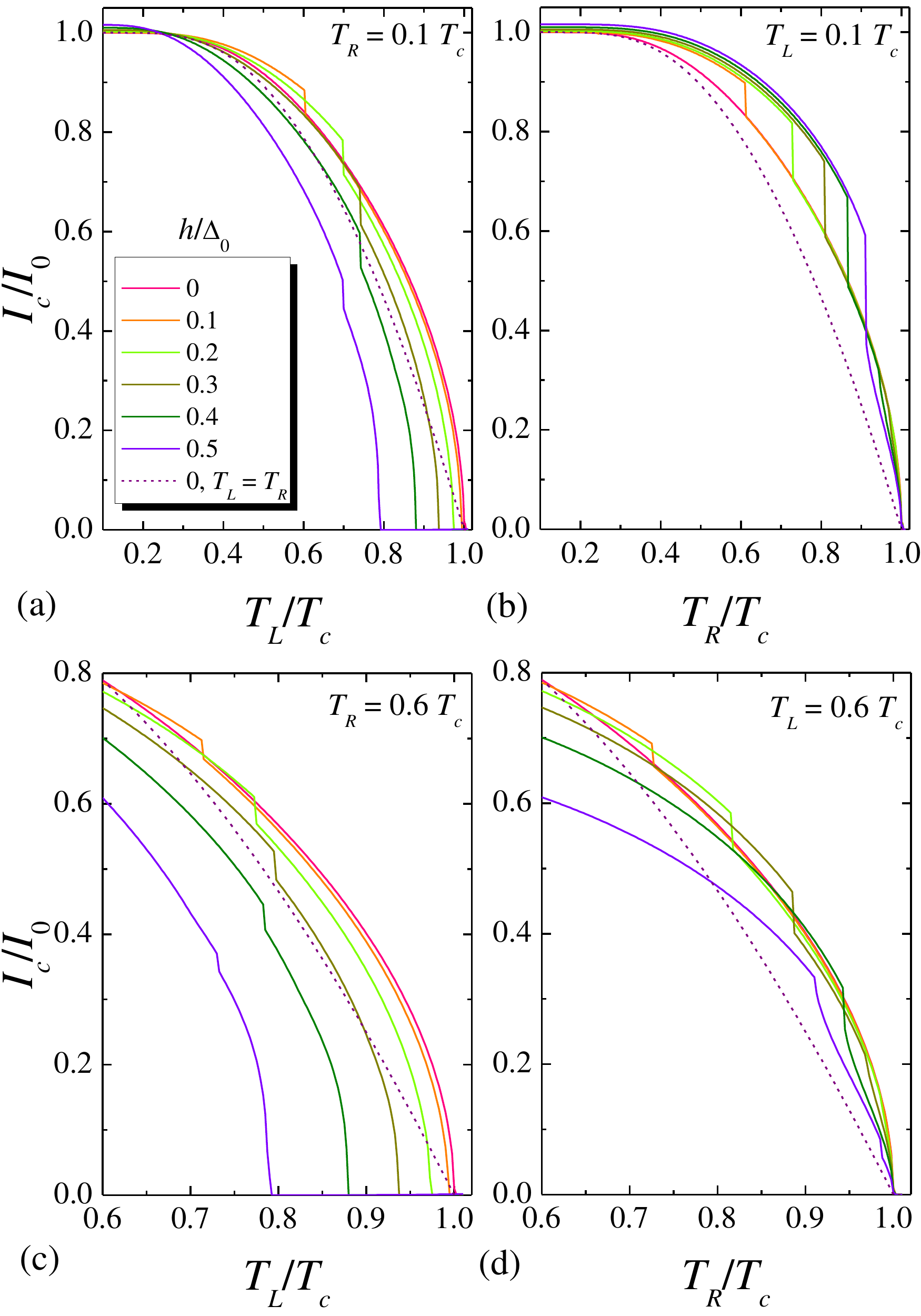}
\caption{(Color online) (a) Critical current $I_c$ versus $T_L$ calculated for selected values of the exchange field $h$ at $T_R=0.1 T_c$. 
(b) $I_c$ versus $T_R$ calculated at $T_L=0.1 T_c$ for the same values of $h$ as in panel (a). (c) The same as in panel (a) calculated for $T_R=0.6T_c$. (d) The same as in panel (b) calculated for $T_L=0.6T_c$. In all panels the  dashed line shows the critical current  for $T_L=T_R$ and in the absence of an exchange field.}\label{fig3}
\end{center}
\end{figure}

In an experimental situation it is somewhat difficult to tune \emph{in-situ} the exchange field present in FI-S layer, and  to verify the $I_c(h)$ dependence as displayed in Fig. \ref{fig2}. 
 However, there is a simpler alternative way to proceed and to demonstrate these effects. Toward this end we propose a possible experimental setup sketched in Fig. \ref{fig1}(b). The structure can be realized through standard lithographic techniques, and consists of a generic   FI-S-I-S Josephson junction where the two S$_\text{L}$ and S$_\text{R}$ electrodes are connected to additional superconducting (e.g., made of aluminum) probes through oxide barriers so to realize normal metal-insulator-superconductor (NIS) tunnel junctions. 
The NIS junctions are used to \emph{heat} selectively the S$_\text{L}$ or  S$_\text{R}$ electrode as well as to perform accurate electron thermometry \cite{giazottormp}.
 Therefore, instead of varying the exchange field in the Josephson weak-link, one could now hold one of  the junction electrodes at a fixed temperature  and vary the temperature of the other lead while recording the current versus voltage characteristics under conditions of a temperature bias \cite{tirelli2008,savin2004,morpurgo1998,courtois2008,roddaro2011}. 
 In this context, the electric current can be led through the whole structure via suitable outer superconducting electrodes allowing good electric contact, but providing  the required thermal insulation necessary for thermally biasing the Josephson junction.
In addition, the tunnel probes enable  to determine independently the energy gaps in the two superconducting electrodes through differential conductance measurements. 
Moreover, from the materials side, ferromagnetic insulators such as EuO or EuS \cite{Miao,Moodera90,Moodera08} combined with superconducting aluminum could be suitable candidates in light of a realistic implementation of the structure.

The critical current behavior under thermal-bias conditions is displayed in Fig. \ref{fig3} where,  in panels (a) and (c),  $T_R$ is held at $0.1T_c$ and $0.6T_c$, respectively, and $T_L$ varies. 
Similarly, in panels (b) and (d) we keep $T_L$ constant at $0.1T_c$ and vary $T_R$.  
It clearly appears that the $I_c(T)$ curves drastically deviates form those obtained at equilibrium,  i.e., for $T_L=T_R$ [dotted lines in Fig. \ref{fig3})].  
If we keep a constant temperature $0.1T_{c}$ in one of the electrodes [see Fig. \ref{fig3}(a) and (b)], and change the temperature of the other it follows that, for low enough temperatures, $I_c$ gets larger by increasing the magnitude of the exchange field. 
This corresponds to the enhancement discussed in Fig. \ref{fig2}. 
By further increasing  the temperature of one of the electrodes leads to a critical current decrease. 
Notably, $I_c$ exhibits a sharp jump at those temperatures such that the condition expressed by Eq. (\ref{condition}) holds. 
This is a striking effect which can  provide, from the experimental side, evidence of the supercurrent enhancement discussed above.
Yet, it can be used as well to determine the value of the effective exchange field induced in the superconductor placed in direct contact with the FI layer.  
It is remarkable that these features can be also observed, although reduced in amplitude,  in the high-temperature regime [see Figs. \ref{fig3}(c) and (d)]. 
We emphasize that the effect here discussed is much more pronounced when the left electrode (i.e., the one with the FI layer) is kept at a low temperature, and one varies $T_R$.
This is simple to understand, since  a superconductor with  a spin-splitting field is more sensitive to a temperature variation: the larger the exchange field the faster one get suppression of superconductivity by enhancing the temperature.

In conclusion, we have shown  that the critical current $I_c$  of a FI-S-I-S Josephson junction is drastically modified by the presence of the exchange field induced in one of the electrodes  from the contact with a ferromagnetic insulator. 
In particular, we have demonstrated  that  the Josephson coupling is strengthened by  the presence of the exchange field and therefore the $I_c$ amplitude is enhanced. The enhancement becomes more pronounced upon the application of a temperature bias across the junction. 
In such a case we predict a  change of the $I_c(T)$ curve with respect to the equilibrium situation which now shows a jump occurring when the difference of the superconducting gaps equals the amplitude of the exchange field. 
This behavior can be measured through standard techniques as we have discussed  for a realistic   experimental setup. 
Our predictions on Josephson junctions with ferromagnetic insulators are of great relevance since they constitute the building blocks of  recently proposed nanodevices for spintronics\cite{Nazarov2002,GB2013c} and coherent spin caloritronics\cite{GB2013a,GB2013b}.

The work of F.S.B was supported by the Spanish Ministry of Economy and Competitiveness under Project FIS2011-28851-C02-02. F.G. acknowledges the Italian Ministry of Defense through the PNRM project "Terasuper", and the Marie Curie Initial Training Action (ITN) Q-NET 264034 for partial financial support. F.S.B thanks Prof. Martin Holthaus and his group for their kind hospitality at the Physics Institute of the Oldenburg University.


\begin{thebibliography}{99}
\bibitem{gamma} Throughout the paper we set $\Gamma =10^{-6} \Delta_0$ for the numerical evaluation of $I_c$.
\bibitem{note2} We stress that the \emph{enhancement} of $I_c$ as a function of $h$  in a F-S-I-S-F junctions only takes place if the magnetizations of the F layers are  oriented in the \emph{antiparallel} configuration. In the parallel configuration $I_c$ is a monotonically decreasing function of $h$.
\end{thebibliography}
  \end{document}